\documentclass[12pt,dvips]{article}
\usepackage{amssymb} % for \Bbb or \mathbb, respectively
\usepackage{epsfig}

\newcommand{\RE}{\mathbb{R}}
\newcommand{\be}{\begin{equation}}
\newcommand{\ee}{\end{equation}}
\newcommand{\psibarpsi}{\bar{\psi}\psi}

\begin{document}
%%%%%%%%%%%%%%%%%%%%%%%%%%%%%%%%%%%%%%%%%%%%%%%%%%%%%%%%%%%%%%%%%%%%%%%%%%%

\thispagestyle{empty}

\date{\today}
\title{
\vspace{-5.0cm}
\begin{flushright}
{\normalsize UNIGRAZ-}\\
{\normalsize UTP-09-04-98}\\
%{\normalsize hep-lat/9804012}\\
\end{flushright}
\vspace*{2.5cm}
Chiral properties of the fixed point action\\
of the Schwinger model\thanks{Supported by Fonds 
zur F\"orderung der Wissenschaftlichen Forschung 
in \"Osterreich, Project P11502-PHY.} }
\author
{\bf F. Farchioni\footnote{E-mail: fmf@physik.kfunigraz.ac.at}, C. B. Lang and M. Wohlgenannt \\  \\
Institut f\"ur Theoretische Physik,\\
Universit\"at Graz, A-8010 Graz, AUSTRIA}
\maketitle
\begin{abstract}
We study the spectrum properties for a recently constructed 
fixed point lattice Dirac operator. We also consider the 
problem of the extraction
of the fermion condensate, both by direct computation, and through the
Banks-Casher formula by analyzing the density of eigenvalues of a
redefined antihermitean lattice Dirac operator.
\end{abstract}


\newpage

\section{Introduction}

One of the most annoying drawbacks of the lattice discretization of a
fermionic theory is the impossibility\footnote{With the precondition
that very general properties of the continuum theory, like locality,
are preserved.} \cite{NiNi} to avoid the explicit breaking of
the chiral symmetry.  In the case of a simple discretization, the
Wilson action, the breaking of the symmetry is so bad that no trace of
the chiral properties of the continuum theory is kept in the lattice
theory. This is of particular relevance for attacking QCD by lattice
techniques.  At the classical level, the chiral properties of the
zero-modes of the Dirac operator are lost, and the Atiyah-Singer
theorem of the continuum theory \cite{AtSi71} has no
strict correspondence on the lattice. At the quantum level, the
explicit breaking of the symmetry induces an additive renormalization
of the quark mass, the chiral limit being attained through a fine
tuning of the bare parameters; moreover, chiral currents undergo a
finite renormalization, mixing among operators with different (nominal)
chirality occurs, and no order parameter for the spontaneous breaking
of the chiral symmetry can be defined in a natural way. `Accidental'
zero-modes (related to the so-called `exceptional configurations') are
causing troubles in Monte Carlo simulations.

In an early paper~\cite{GiWi82} of lattice quantum field theory,
Ginsparg and Wilson already gave the key to make definite what is meant
by chiral limit in the framework of a lattice theory breaking
explicitly chiral symmetry, providing a general condition -- we will
refer to it as to the Ginsparg-Wilson Condition (GWC) -- for the
fermion matrix of the lattice theory. They showed in particular that
any action fulfilling the GWC reproduces the correct triangular
anomaly.  The condition was found by requiring that the lattice action
is obtained as the fixed point (FP) action of a block-spin
transformation (BST), in the universality class of a continuum
chiral-symmetric action.  In this way, the breaking of the symmetry is
introduced in the lattice action by the BST itself and not `by hand'.
Recently it has been shown \cite{Ne98} that the GWC can be solved in a
rather independent approach, namely in the so-called overlap formalism,
and the corresponding symmetry has been identified \cite{Lu98}.

Ginsparg and Wilson's observation had no practical applications until
the technology for the construction of FP actions of lattice gauge
theory was improved following Hasenfratz and Niedermayer's ideas
\cite{HaNi94,DeHaHa95,BiWi96}.  In a recent series of papers it has
been shown that the GWC is a sufficient condition for the restoration
of the main features of the continuum (symmetric) theory; in
particular, at the classical level, the Atiyah-Singer theorem finds
correspondence on the lattice~\cite{Ha98c,HaLaNi98}, excluding the
possibility of `accidental' zero-modes; at the quantum level, no fine
tuning, mixing and current renormalization occur, and a natural
definition for an order parameter of the spontaneous breaking of the
chiral symmetry is possible~\cite{Ha98a}.

Monte Carlo calculations require actions with finite number of
couplings, while the FP action is extended over all distances (even
with exponential damping of the couplings). It is therefore clear that
only a {\em parametrized} form of the FP action is of practical
relevance. The point to be checked is to what extent such an
approximation of the FP action is able to reproduce the nice properties
of the `ideal' FP action.  The construction of the FP action of QCD is
- because of technical problems -- yet still far away (for some
pioneering attempts, see \cite{De98a}), and meanwhile we try to
practice with toy models.

Recently, a parameterization of the FP action of the Schwinger model for
the non-overlapping BST (noBST) (we refer to this parameterization as to
$pA_{FP}$ subsequently, while to the `ideal' FP action simply as to
$A_{FP}$) was found \cite{LaPa98} in a Monte Carlo approach. The
fermionic action is parametrized in terms of bilinear expressions in
the fermion fields, connected by paths made up of
compact gauge link variables, whereas the gauge action was the
non-compact simple fixed point action. Here we give a brief account of
a study of the spectral properties of this Dirac operator, concerning
the numerical verification of some of the chiral properties of the FP
action in the case of the $pA_{FP}$.  A more detailed report will be
given elsewhere. We focus our attention on the
spectral properties of the FP Dirac operator, verifying the lattice
Atiyah-Singer theorem.  A first verification of this theorem was
accomplished in a different approach in \cite{FaLa98}.  Moreover, we
address the problem of the extraction of the fermion condensate, both
from the direct computation and through the Banks-Casher
formula~\cite{BaCa80}.

\section{Chiral Properties of the FP Action}

The GWC for the fermion matrix $h_{x,x^{\prime}}$ for massless fermions
reads
\be
\frac{1}{2}\left\{\, h_{x,x^{\prime}},\gamma^5\,\right\}\;=\;\left(\,
h\,\gamma^5\,R\,h\,\right)_{x,x^{\prime}}\;\;,
\label{eq:gwcg}
\ee
where $R_{x,x^{\prime}}$ is a local matrix in coordinate space,
i.e. whose matrix elements vanish exponentially with the distance.
Any FP action of a BST satisfies the GWC: $R_{x,x^{\prime}}$ is fixed
by the average matrix of the fermion BST, and it can be in general
assumed to be trivial in Dirac space.

In the case of the non-overlapping BST (noBST) 
considered in \cite{LaPa98}, one has
$R_{x,x^{\prime}}=\frac{1}{2}\delta_{x,x^{\prime}}$, and the GWC
assumes the elegant form
\be
h\,+\,h^{\dagger}\:=\:h^{\dagger}\,h\:=\:h\,h^{\dagger}\;\;.
\label{eq:gwc}
\ee
where also the hermiticity property $h^{\dagger}=\gamma^5h\gamma^5$
(fulfilled by the FP action in this study as well as by other actions)
has been taken into account.

\subsection{The spectrum of the Dirac operator}

Eq.~(\ref{eq:gwc}) implies non trivial properties of the spectrum of
the Dirac operator, namely the fermion matrix $h$:
\begin{itemize}
\item [i.] $[h,\, h^{\dagger}]\:=\:0$, i.e. $h$ is a normal operator; as
a consequence, its eigenvectors form a complete orthonormal set.
\item [ii.] The spectrum lies on a unitary circle
in the complex plane centered at $(1,0)$.
\item [iii.] The property (i), together with
the hermiticity property of the fermion matrix, implies (denoting with
$v_{\lambda}$ an eigenvector of $h$ with eigenvalue $\lambda$):
\begin{eqnarray}\label{definitechirality}
\gamma^5\,v_{\lambda}&=&v_{\lambda^*}\quad,\qquad\mbox{if}\quad\lambda\not=\lambda^*\nonumber
\\
\gamma^5\,v_{\lambda}&=&\pm\,
v_{\lambda}\;\;,\qquad\mbox{if}\quad\lambda=\lambda^*\in\RE \;.
\end{eqnarray}
So, just as in the continuum the eigenvectors of complex-conjugated
eigenvalues form chirality doublets; moreover all real-modes have
definite chirality. (For a general FP action this property holds
only for the zero-modes.)
\end{itemize}

It was shown in \cite{HaLaNi98} that (\ref{eq:gwcg}) ensures the
existence of a lattice version of the Atiyah-Singer theorem,
which can be stated in the form:
\be
Q_{FP}\:=\:-\sum_{\{v_0\}}\:(v_0,\gamma^5 v_0)\;\;,
\label{eq:AST}
\ee
where $Q_{FP}$ is the fixed point topological charge 
\cite{BlBuHa,DeHaZh96} of the background gauge configuration. 
$Q_{FP}$ provides a lattice definition of the 
topological charge of a gauge configuration which depends on the RG blocking
procedure for the gauge sector.

\subsection{The fermion condensate}

In \cite{Ha98c}, a subtraction procedure for the fermion condensate -- 
inspired by (\ref{eq:gwcg}) -- was proposed:
\be
\langle\psibarpsi\rangle_{\rm sub}\:=\:-1/V\,\langle\, {\rm
tr}(h^{-1}-R)\,\rangle_{\rm gauge}\;\;,
\label{eq:subfc}
\ee
where $V$ is the (finite) space-time volume.  Because
of~(\ref{eq:gwcg}) the quantity in brackets of the r.h.s of the above
equation vanishes, except when a zero-mode of $h$ occurs, in which case
a regulator-mass $\mu$ must be introduced, $h\to h_\mu=h+\mu$.  For the
number of flavors $n_f>1$, the contribution of the zero-modes to the
gluon average vanishes  when $\mu\rightarrow 0$ because of the damping
effect of the fermion determinant, and we conclude that the
finite-volume subtracted fermion condensate vanishes in the chiral
limit \cite{Ha98c}.  The case $n_f=1$ is peculiar since
$\langle\psibarpsi\rangle_{\rm sub}\not=0$ even in a finite volume.
The configurations responsible in this case for the non-zero fermion
condensate in  a finite volume are those from the $|Q|=1$ sector;
indeed, if $h$ has just one zero-mode -- which is possible because of
the (lattice) Atiyah-Singer theorem only for $|Q|=1$ -- the quantity
${\rm tr}\,h^{-1}({\rm det}\,h)$ has a non-zero limit when
$\mu\rightarrow 0$.  The effect of the subtraction is in general (i.e.
for any $n_f$) just to remove the spurious contribution of the $Q=0$
sector introduced by the explicit breaking of the chiral symmetry of
the FP action.

The situation in the infinite volume, i.e. when the fermion condensate
is obtained through the sequence of limits $\lim_{\mu\rightarrow
0}\lim_{V \rightarrow \infty}$, is somehow different:  in this case the
role of the zero-modes is irrelevant\footnote{We are grateful to P.
Hasenfratz for having driven our attention on this point.}, the
quasi-zero-modes being responsible for the non-zero fermion condensate
as the Banks-Casher formula shows.

We observe that (\ref{eq:subfc}) may be rewritten in the
form
\be\label{directdef}
\langle\psibarpsi\rangle_{\rm sub}\:=\:
-1/V\,\langle\, {\rm tr}(\tilde{h}^{-1})\,\rangle_{\rm gauge}\;\;,\;\;\;\;
\tilde{h}\:=\: h\, (1-R\,h)^{-1}\;\;.
\ee
The redefined fermion matrix  $\tilde{h}$ has the non trivial properties:
\be
\tilde{h}^{\dagger}\,=\,-\tilde{h}\;\;,\;\;\;\;\{\tilde{h},\gamma^5\}\,=\,0\;\;,
\ee
from which it follows that $\tilde{h}$ has a purely imaginary
spectrum,
\be
\tilde{h}\,v=-i\,\tilde{\lambda}\,v\;\;,\;\;\;\;\tilde{\lambda}\in\RE\,;
\ee
of course, the zero-modes of $h$ coincide with those of $\tilde{h}$.

The replacement $\lambda\rightarrow\tilde{\lambda}$ allows in a sense
to reconstruct the `chirally invariant information' of the 
spectrum of $h$. One can show that the spectral density of $\tilde{h}$,
$\tilde{\rho}(\tilde{\lambda})$ complies
(in the limit: $\lim_{ a\rightarrow 0}\lim_{V\rightarrow\infty}$) with the
the Banks-Casher formula for the subtracted fermion condensate:
\be\label{BanksCasher}
\langle\psibarpsi\rangle_{\rm sub}\:=\:-\pi\,\tilde{\rho}(0)\;\;.
\ee
In the case of the noBST, one has
\be\label{projection}
\tilde{\lambda}=\lambda\,\left(1-\frac{\lambda}{2}\right)^{-1}
\ee
and the spectrum of $\tilde{h}$ is obtained by mapping
the spectrum of $h$ (which, we recall, lies on a unitary
circle in the complex plane) onto the imaginary axis 
by the stereographic projection.

\section{Numerical Results: the Schwinger Model}

The parametrized FP action $pA_{FP}$ of the Schwinger model in 
\cite{LaPa98} was obtained in the non-compact formulation 
for the pure-gauge sector.  
A first difficulty arises in this respect, since it is not possible to
give a lattice definition of the topological charge in terms of
the non-compact gauge variables. The most natural definition, i.e.
$1/2\pi\sum_x F_{12}(x)$ vanishes identically on the torus. 
A definition in terms of compact variables is instead available, 
the so-called `geometrical' charge:
\be
Q_{G}\;=\;\frac{1}{2\pi}\,\sum_x\,{\rm Im\, ln}(U_{12}(x))\; .
\label{geocha}
\ee

Measuring the topological charge after the compactification of the 
gauge variables according to (\ref{geocha}) reveals however
an unnatural suppression of non-zero values even at moderate $\beta$'s,
due to the non-compact nature of the gauge action.

Ideally, one should work with the FP action of the compact
theory, the corresponding FP topological charge operator
coinciding\footnote{Except for a set of measure zero,
when $U_{12}(x)=-1$ for some $x$, 
in which case the prescription (\ref{geocha}) is not well-defined.}
in this case with the expression (\ref{geocha}) \cite{FaLa98}.
Since a parameterization of a compact FP action is
not yet available, here, as an approximation, we take the
parametrized FP fermion matrix of \cite{LaPa98} (which is  compact
by construction) and simply replace the original action of the
pure-gauge sector with the compact Wilson action.  Of course this
approximation is expected to introduce additional deviations from the 
behaviour of the ideal FP action $A_{FP}$ when topological fluctuations 
are important in the statistical ensemble.

The fermion part of $pA_{FP}$ has the form 
\begin{equation} \label{fermfit} %\begin{array}{l}
\bar{\psi}\,h_p(U) \,\psi =
\sum_{i=0}^3\sum_{x\, , f}\,  \rho_i(f)\, 
\bar{\psi}(x)\, \sigma_i\, U(x,f)\, \psi(x+\delta f)\;.
\end{equation}
Here $h_p(U)$ is the parametrized lattice Dirac operator, $f$ denotes a closed
loop through $x$ or a path from the lattice site $x$ to $x+\delta f$
(distance vector $\delta f$) and $U(x,f)$ is the parallel transporter
along this path. The $\sigma_i$-matrices denote the Pauli matrices for
$i= 1,2,3$ and the unit matrix for $i=0$. We make the identification:
$\gamma^{0}=\sigma_1$, $\gamma^{1}=\sigma_2$, $\gamma^5=\sigma_3$.
The action obeys the usual
symmetries as discussed in \cite{LaPa98}; altogether it has 429 terms per site.

We constructed gauge field configurations according to (a) the compact
Wilson action $S_W$ and (b) the non-compact action
$S_{NC}=(\beta/2)\sum_x F_{12}^2(x)$. In the case (a) the configurations
were generated with a Metropolis Monte  Carlo update, separating
configurations by a number of updates of twice the size of the
integrated autocorrelation length for the (geometric) topological
charge, which scales with $\beta$ exponentially (e.g. approximately
$\tau_Q\simeq \exp{(1.67\,\beta-3)}$ for $\beta>2$ on $16^2$ lattices;
for comparable observations see \cite{ElBu}).  The measured
distributions for the topological charge agree with the observations in
\cite{GaHiLa97}. In situation (b) the configurations were generated
completely independently according to the Gaussian measure (respecting
the gauge d.f.).

For each of those gauge field configurations we analyzed the Dirac
operator eigenvalues (and sometimes eigenvectors) and the determinant.
So most of the results presented may be considered quenched results,
although some of the observables have been weighted with the
determinant.  The latter mentioned numbers are for the full, dynamical
system, although maybe plagued by the expectation value of the
determinant in the denominator -- as we will discuss below.

\subsection{The spectrum}

\begin{figure}[t]
\begin{center}
\epsfig{file=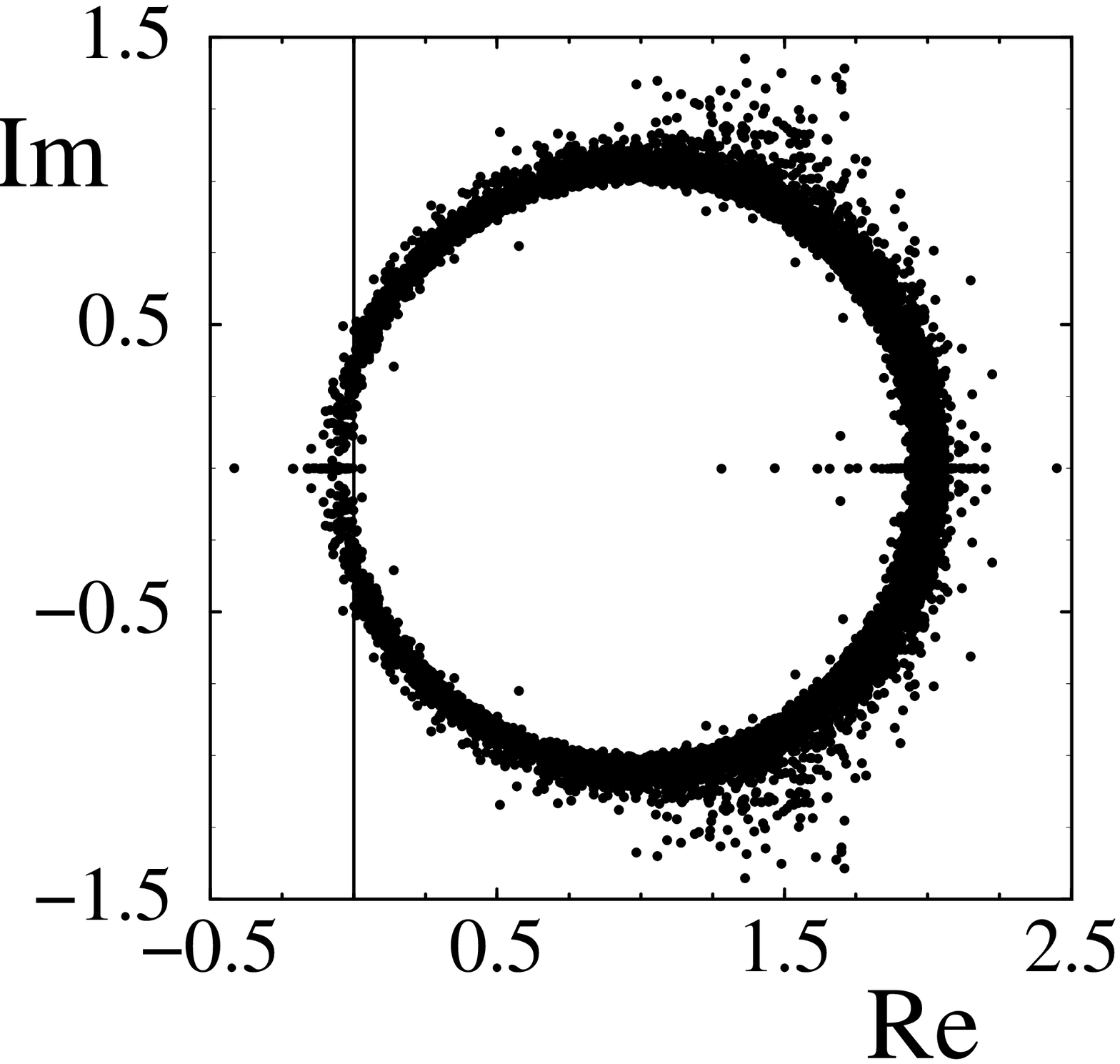,width=4.2cm,angle=-90}
\epsfig{file=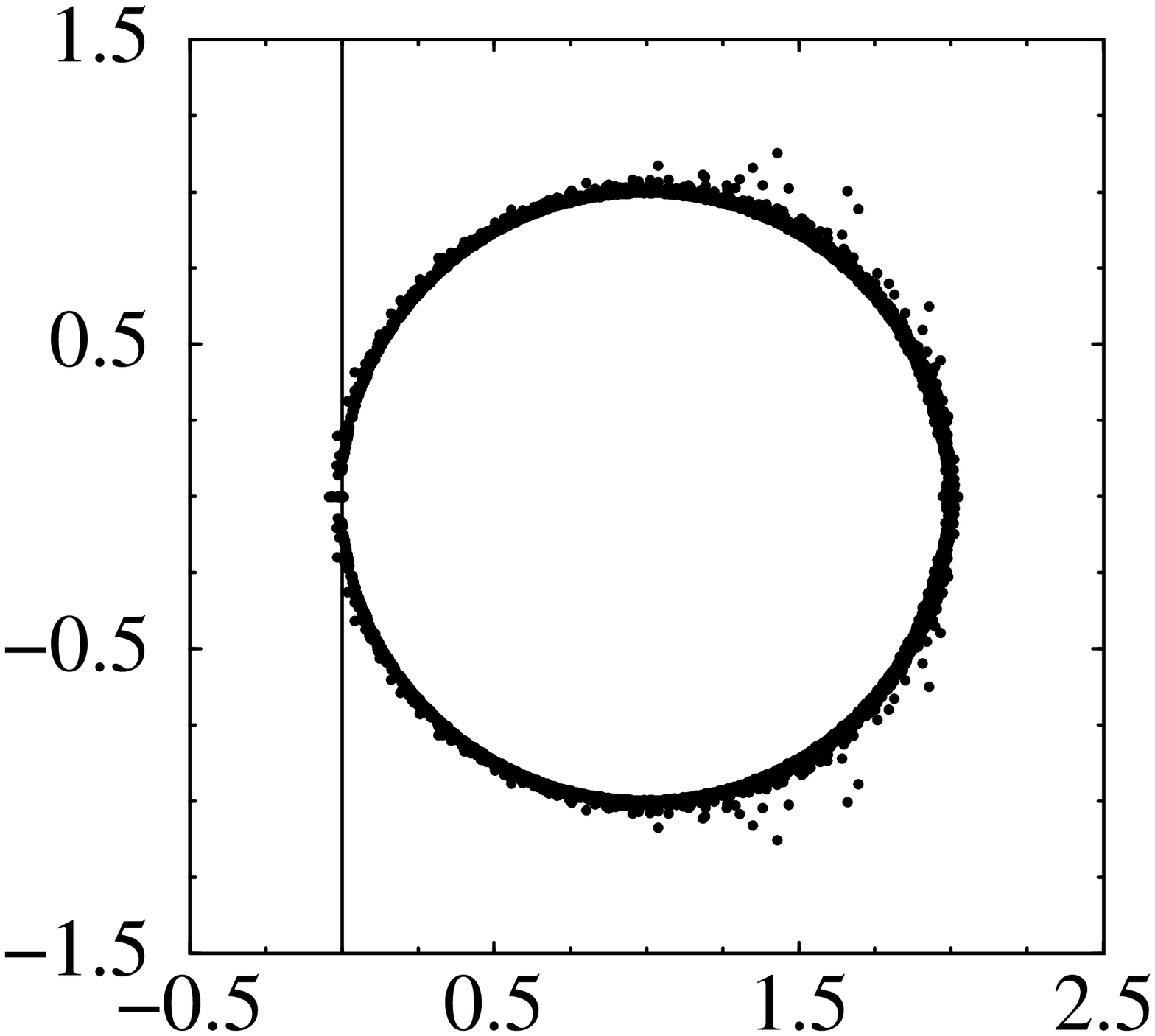,width=4.2cm,angle=-90}
\epsfig{file=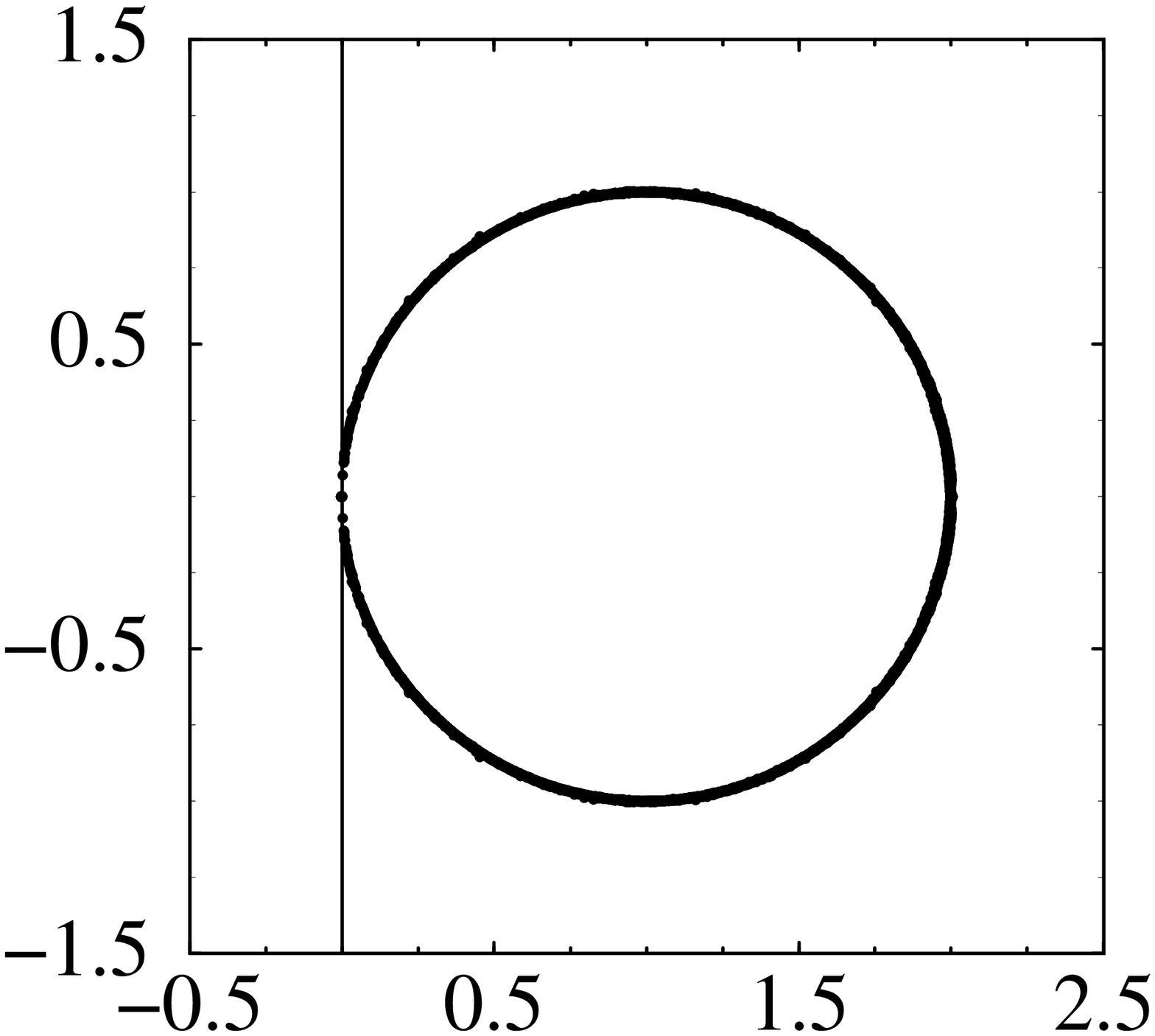,width=4.2cm,angle=-90}\\
\epsfig{file=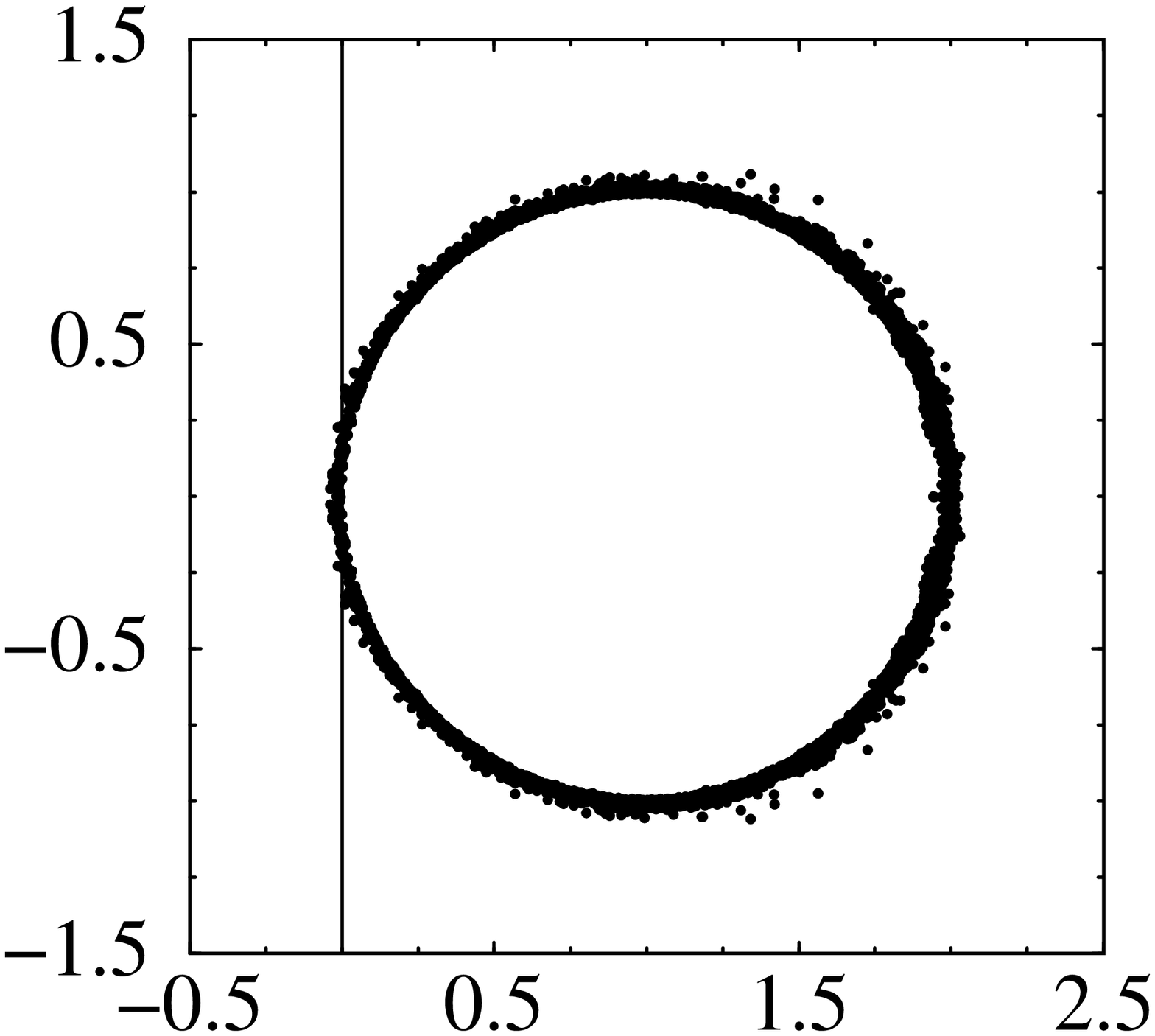,width=4.2cm,angle=-90}
\epsfig{file=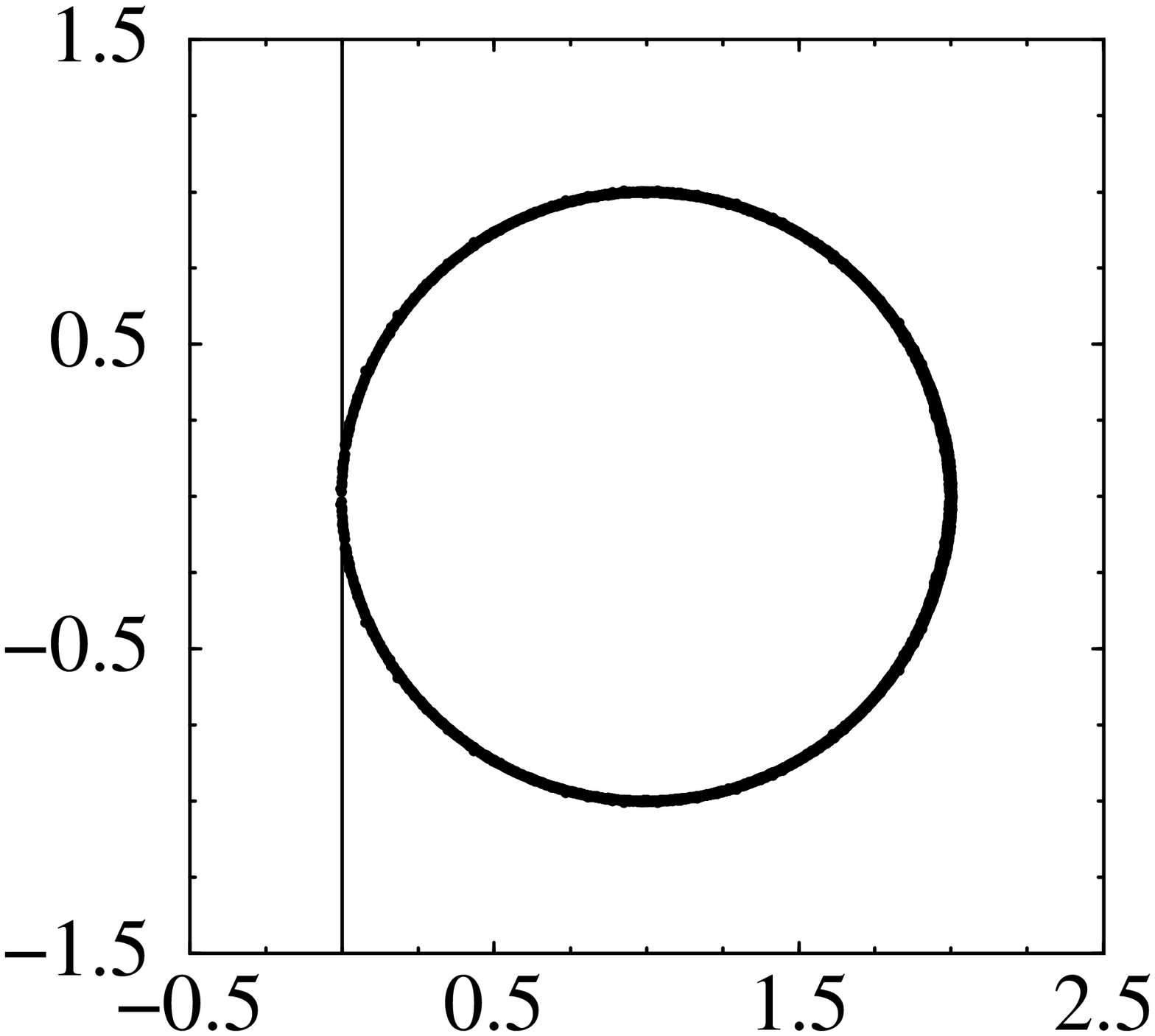,width=4.2cm, angle=-90}
\epsfig{file=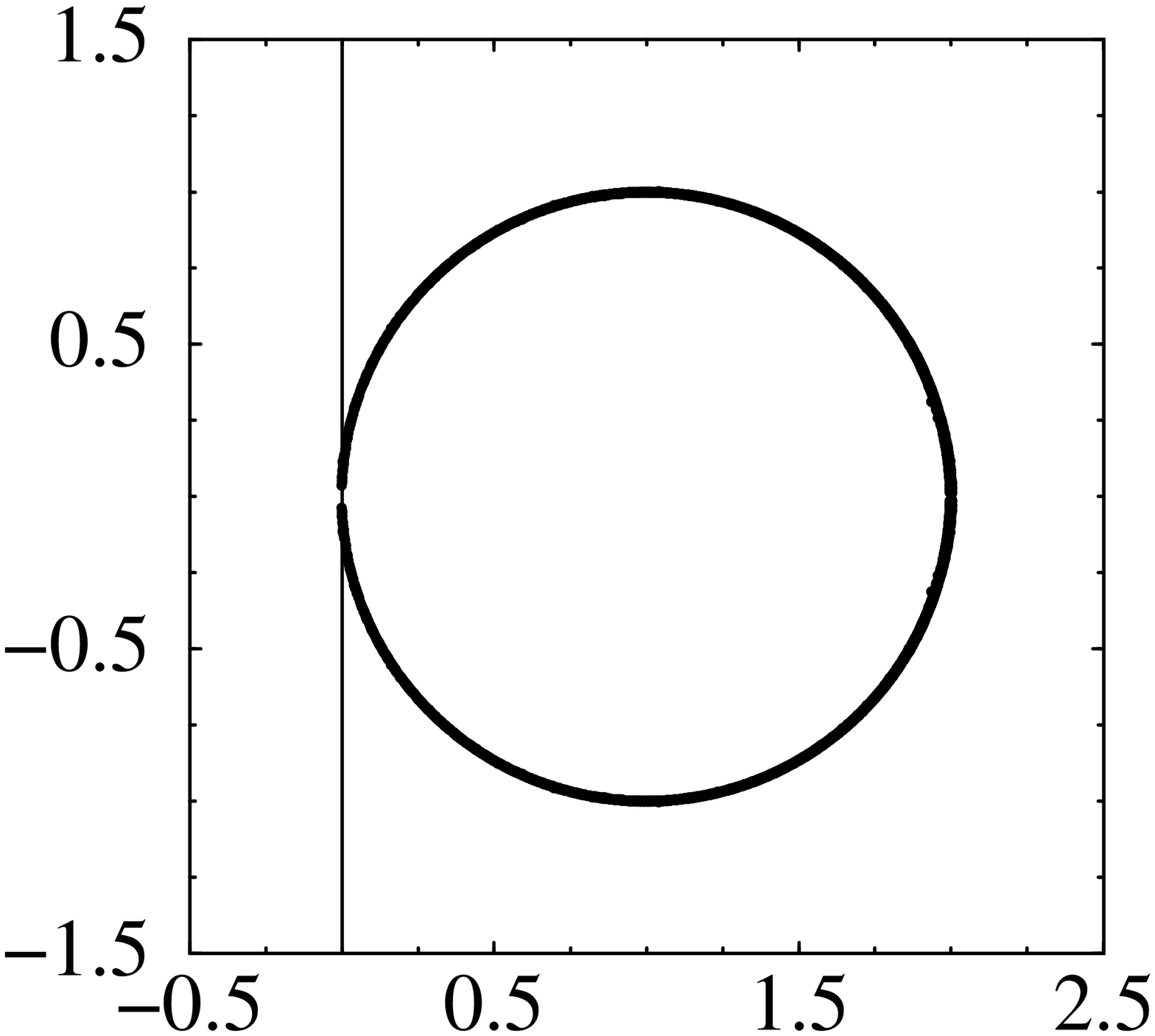,width=4.2cm,angle=-90}
\end{center}
\caption{\label{fig1}
Eigenvalues of the lattice Dirac operator (\ref{fermfit}) 
at the values of
$\beta=2, 4, 6$ (from left to right), on $16^2$-lattices and sampled over 
25 gauge configurations each, according to the compact (upper row)
or non-compact (lower row) gauge action.}
\end{figure}

Fig. \ref{fig1} shows the spectrum of the studied lattice
Dirac operator (\ref{fermfit}) collectively for 25 gauge configurations
(uncorrelated to the amount discussed earlier; we collected data for several 
hundred such configurations, but do not plot them in order to prevent
the figure files becoming too large. The overall behaviour is well 
represented in the figures). At larger values of
$\beta$ we notice excellent agreement with the expected circular shape
for $A_{FP}$. For smaller $\beta$ the spectrum roughens, although
the circular shape generally is kept.

One has to keep in mind that the numerically determined parametrized
Dirac operator (\ref{fermfit}) is truncated and optimized on a finite
sample of (non-compact) gauge field configurations. Thus it is not
surprising, that the spectral shape is closer to the optimum for the
configurations sampled according the non-compact gauge action in Fig.
\ref{fig1}.

Subsequently we discuss (except when explicitly stated differently) our
results for the compact gauge action, for $n_f=1$ and lattice volume
$16^2$.

We note, that there are configurations with real eigenvalues. We
checked the eigenvectors for those and confirm that these modes have
definite chirality, indeed, as expected from
(\ref{definitechirality}).  Also, we can clearly distinguish the real
values around zero from those around 2 (right-hand part of the
spectrum). In fact each real quasi-zero-mode has three real
`doubler'-partners at the other end of the spectrum. The chiralities of
the partners add up to the opposite value of the chirality of the
quasi-zero-mode(s).

We may identify these real eigenvalues (around zero) with zero-modes
and relate their number $n_0$ with the geometrically (i.e. from the
gauge field configuration) defined topological charge $Q_G$.  We find
agreement in the following sense: The ratio of the number of
configurations, where these numbers coincide over those, where they do
not, approaches unity in the limit $\beta\to\infty$. In fact, the
approach is faster than that observed in the situation of the Wilson
Dirac-operator \cite{GaHiLa97}.  We find that ratio to be 0.887 , 0.999
and 1.000 at $\beta=2$, 4 and 6, respectively.

Whereas $A_{FP}$ has no real modes at negative values, here we do
observe those, in particular towards smaller $\beta$. This is expected
due to the general roughening (cf. Fig. \ref{fig1}) and is equivalent
to the (albeit more frequent) occurrence of `exceptional'
configurations for the Wilson fermion action already below $\kappa_c$.

\begin{figure}[ht]
\begin{center}
\epsfig{file=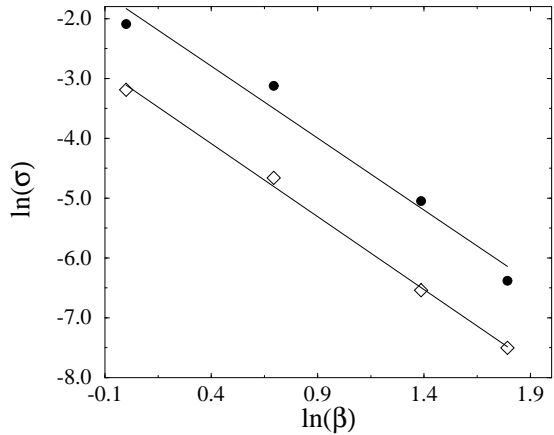,angle=-90,width=8cm}
\end{center}
\caption{\label{fig2}
Scaling behaviour of the distribution width $\sigma$ for the spectra 
observed in Fig. 1 a--c (full circles, compact gauge action) 
and Fig. 1 d--f (diamonds, non-compact gauge action) as discussed 
in the text. The lines are the result of a linear fit.}
\end{figure}

In order to estimate the scaling behaviour of the deviations of the
spectrum from the ideal circular shape, we defined a mean deviation
$|\lambda-1|$ from the unit circle in the region close to $\lambda=0$,
in an angular window of $|\mbox{arg}(1-\lambda)|<\pi/4$. Fig.
\ref{fig2} shows the behaviour of the average width (standard
deviation) $\sigma$ of that distribution with regard to $\beta$. The
log-log plot demonstrates a behaviour of $\sigma$ $\propto
1/\beta^{2.41}\simeq a^5$.  The parametrized action $pA_{FP}$ is
truncated in a finite range (7x7 in our case).  Heuristically this
implies an error for the eigenvalues, considered as dimension-one
gauge-invariant operators of the gauge field, in the form of some
operator of higher dimension $k$.  From the observed deviation we
estimate an effective value $k\simeq 5$. Since similar behaviour
(differing just in an overall multiplicative factor) is observed for
both types of gauge action, we think that we may justify the observed
deviation by the truncation.

\subsection{The fermion condensate}

\paragraph{Spectral density:} The spectral density for the projected
eigenvalues $\tilde \lambda$ (see (\ref{projection})),
$\tilde{\rho}(\tilde{\lambda})$, has been calculated both in the
quenched and unquenched situation.  In the unquenched case, the
individual configurations are weighted by the value of the determinant,
which amounts to sampling in the presence of dynamical fermions.  The
fluctuations are in this case higher, since one has to divide by the
sum over determinant values, and as is well known this involves
cancellation of large terms and therefore substantial statistical
errors.

\begin{figure}[t]
\begin{center}
\epsfig{file=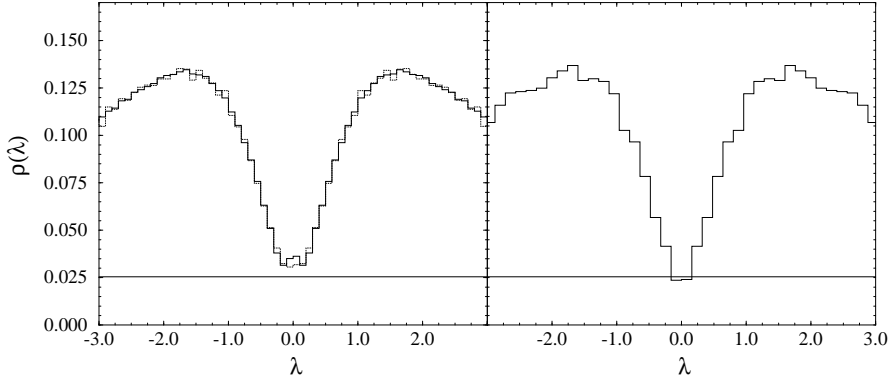,angle=-90,width=12cm}
\end{center}
\caption{\label{fig3}
Density of the eigenvalues (projected according (\ref{projection}) from
the close-to-circular positions to the close-to-imaginary axis
positions) near zero; $\beta=4$, volume $16^2$. The horizontal line
gives the theoretical expectation from the Banks-Casher formula.  Left:
Quenched histogram and histogram without zero-modes (dotted).  Right:
The histogram weighted by the determinant, i.e.  the unquenched
distribution.}
\end{figure}

The results are displayed in Fig. \ref{fig3} for a sample of 2000
configurations.  In the quenched case we observe in the central part of
the spectrum a peak (smoothed by the binning procedure) produced by the
(quasi-) zero-modes. It is removed when they are discarded from the
sample (cf. Fig. \ref{fig3}). This peak is absent in the unquenched
situation as expected:  The zero-modes are suppressed due to the
determinant. Apart from this, the unquenching does not change the main
features of the distribution, only introducing higher fluctuations.
For both, the so `cleaned' distribution as well as for the full,
unquenched case, we observe good agreement of $\rho(0)$ with the
theoretically expected value for the condensate at infinite volume
$\langle\psibarpsi\rangle= -{\rm e}^{\gamma}/(2\pi\sqrt{\beta\pi})$
through the Banks-Casher relation (\ref{BanksCasher}).  The theoretical
value in the finite volume case (torus) is also available
\cite{SaWi92}; for the physical volume here considered the correction
is less than $4\%$.

Our conclusions from these observations are:
\begin{itemize}
\item The partial unquenching (by omitting the zero-modes)
gives similar results as the complete unquenching
(by including the determinant).
\item The zero-modes produce a substantial finite-volume
effect; indeed their elimination allows us to approximately recover 
the Casher-Banks relation even in the quenched case. 
\end{itemize}

\begin{figure}[t]
\begin{center}
\epsfig{file=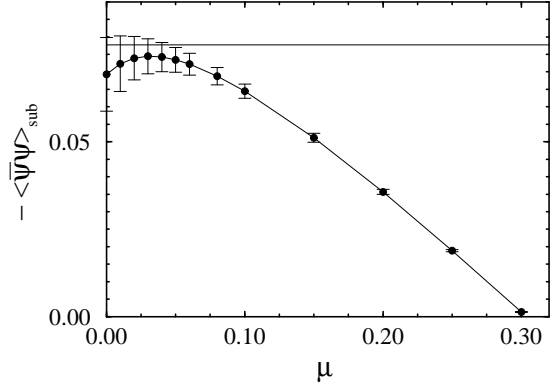,width=8cm}
\end{center}
\caption{\label{fig4}
We show $-\langle\psibarpsi\rangle_{sub,\mu}$ determined from the
same sample of configurations as discussed in Fig. \ref{fig3}.}
\end{figure}

\paragraph{Direct computation:}
Alternatively we may attempt to derive $\langle\psibarpsi\rangle$
directly from ${\rm tr}(\tilde{h}^{-1})$ according 
to (\ref{directdef}).

The standard procedure to get the infinite volume fermion condensate is
to take the succession of limits $\lim_{\mu\rightarrow 0}\:
\lim_{V\rightarrow\infty}$.  We argue that we have indeed accomplished
something equivalent to this in the previous calculation with the
spectral density, via the Banks-Casher formula: For that quantity, the
volume effects are small (after discarding quasi zero-modes), so that
the $V\rightarrow\infty$ limit is effectively attained, and the limit
$\tilde\lambda\rightarrow 0$ replaces the limit $\mu\rightarrow 0$.
Note, that these two limits are complementary, since $\mu$ represents
the {\em real} part of the eigenvalues of $\tilde h$, while
$\tilde\lambda$ the {\em imaginary} part.  One can show \cite{LeSm92}
that the two approaches, the direct computation and the one via the
Banks-Casher formula, are consistent even in a {\em finite} volume, if
$\mu\gg 1/V\langle\bar\psi\psi\rangle$.

First of all, we have checked that, as expected from the discussion of
the previous section, the subtracted fermion condensate
(\ref{directdef}) is zero (within error bars) in the trivial
topological sector. We then studied the {\em full} condensate, i.e.
including also (quasi-) zero-modes. Fig.  \ref{fig4} gives
$-\langle\psibarpsi\rangle_{sub,\mu}$ for the discussed sample of 2000
configuration ($\beta=4$, volume $16^2$) determined for several values
of the regulator mass $\mu$. The values are of course strongly
correlated since each point contains the same set of eigenvalues of
${h}$.  The comparison with the (infinite volume) continuum value shows
good agreement near $\mu=0$.  In fact, it turns out, that the regulator
here is not really important.  Essentially all quasi-zero-modes are not
exact zero modes and so the regularization of the determinant
(vanishing for an exact zero mode) and the inverse eigenvalue
(diverging) is not necessary.

Obviously a better study of the finite volume dependence for the
presented quantities is in order. However, already from the results
presented here we conclude, that -- at least for the $n_f=1$ Schwinger
model -- even approximate fixed point actions provide excellent
possibilities to study chiral properties.

{\bf Acknowledgment:}
We wish to thank Peter Hasenfratz, Ivan Hip and Thomas Pany for 
many helpful discussions.

\end{document}